\documentclass[submission,copyright,creativecommons]{eptcs}
\providecommand{\event}{SNR 2017} % Name of the event you are submitting 
\usepackage{amssymb}
\usepackage{amsmath}

\newtheorem{definition}{Definition}

\newtheorem{theorem}{Theorem}

\title{On the Underapproximation of Reach Sets of Abstract Continuous-Time Systems}
\author{Ievgen Ivanov
\institute{Taras Shevchenko National University of Kyiv
\email{ivanov.eugen@gmail.com}}
}
\def\titlerunning{On the Underapproximation of Reach Sets of Abstract Continuous-Time Systems}
\def\authorrunning{I. Ivanov}
\begin{document}

\maketitle

\begin{abstract}
We consider the problem of proving that each point in a given  set of states (``target set'') can indeed be reached by a given nondeterministic continuous-time dynamical system from some initial state. We consider this problem for abstract continuous-time models that can be concretized as various kinds of continuous and hybrid dynamical systems.

The approach to this problem proposed in this paper is based on finding a suitable superset $S$ of the target set which has the property that each partial trajectory of the system which lies entirely in $S$ either is defined as the initial time moment, or can be locally extended backward in time, or can be locally modified in such a way that the resulting trajectory can be locally extended back in time. 

This reformulation of the problem has a relatively simple logical expression and is convenient for applying various local existence theorems and local dynamics analysis methods to proving reachability which makes it suitable for reasoning about the behavior of continuous and hybrid dynamical systems in proof assistants such as Mizar, Isabelle, etc.
\end{abstract}

\section{Introduction}

Real-time embedded and cyber-physical systems (CPS) \cite{Baheti2011,Lee2011} are important types of artificial systems that combine software and hardware and interact closely with  external   devices and the physical environment. 
One of the  important aspect of such systems is their dynamical behavior which depends on the behavior of the physical environment with which they interact. Normally, the latter behavior of the  environment   can be represented using the modeling notion of a global continuous (real) time, so on a certain level of abstraction it makes sense to model the dynamical behavior of an embedded/cyber-physical system also in terms of a global continuous time.
On other levels, more close to the software, one may wish to focus on the discrete-continuous nature of this behavior and consider alternative time  models that are  convenient in this case, like hybrid time (used in hybrid automata) \cite{Goebeletal2009}, superdense time \cite{Lee2011}, non-global time models, etc. Still continuous-time  models play an important role in the definitions of various classes of discrete-continuous  (hybrid) models  \cite{Goebeletal2009} and  special kinds of hybrid models admit reformulation in the form of certain kinds of  well-known continuous-time  models \cite{Goebeletal2009}, one of the most general among which are differential inclusions \cite{Aubin2009}. To illustrate this generality, ordinary differential equations ($x'(t)=f(x(t),t)$),  differential equations with discontinuous right-hand side ($x'(t)=f(x(t),t)$, where $f$ is discontinuous, under various notions of solution like Carath\'{e}odory solution \cite{Fillipov64}, Filippov solution \cite{Fillipov64}, etc.),   implicit differential equations ($f(t,x(t),x'(t))=0$), differential inequalities ($x'(t) \ge f(x(t),t)$),  certain hybrid dynamical systems \cite{Goebeletal2009}, switched systems \cite{Liberzon2003}, etc. can be reformulated \cite[p. 312]{repovs2013continuous} in the form of a differential inclusion \cite{Aubin2009}  $x'(t) \in F(x(t),t)$, where $F$ is a function from a subset of $\mathbb{R}^{n+1}$ to $2^{\mathbb{R}^n}$. (Note, however, that  a differential inclusion admits different notions of solution, e.g. ordinary solution \cite{Bacciotti05generalizedsolutions,Fillipov64}, classical solution \cite{Bacciotti05generalizedsolutions}, $g$-solution \cite{Bacciotti05generalizedsolutions}, etc., and in the case of the mentioned reformulation, different notions of solutions for an initial  model correspond to different kinds of solutions of the resulting  inclusion).
Despite their generality, differential inclusions cannot be a common mathematical ground suitable for describing all dynamical aspects of embedded systems/CPS.
One of the reasons is that they are concrete models that cannot encompass abstractions of various elements of a system in a convenient way. E.g., they require a specific structure of the state space in which their solutions/trajectories take values ($\mathbb{R}^n$ or more general topological vector spaces). If one uses a differential inclusion to model the  behavior of a whole  CPS, one is forced to assume that the state space of the CPS is e.g. $\mathbb{R}^n$, encode the states of various (software, hardware) layers  from which a CPS consists in real vectors, and encode the description of the behavior of software in derivative constraints.
Such an encoding may be used, if it is necessary for automatic analysis/verification of a system, but it is arguably unsuitable for the purpose of  specifying system behavior and reasoning about or refining system behavior specification.

For such purposes abstract continuous-time dynamical models are desirable (that do not impose restrictions on the state space), that admit refinement to concrete dynamical models (e.g. those that can be described by  state transition systems, differential equations, switched systems, etc.) and have certain general analysis principles and property proof principles that can be concretized after refinement. Such abstract dynamical models can be found, e.g. in various variants of the mathematical systems theory (a list of references can be found in \cite{IvTAMC2014}). 

An abstract dynamical system model usually describes a set of abstract trajectories  which are mappings from a time scale to an abstract state space. The behavior of a system is defined by the set of its possible trajectories. In some cases, a  class of abstract  models can be described by  stating  axiomatically the assumed  properties of  trajectories of its systems.

Following this approach, in \cite{Ivanov2012,IvTAMC2014} we defined a class of abstract dynamical system models which we called Nondeterministic Complete Markovian Systems (NCMS) which can be seen as a  subclass of O. H\'{a}jek solution systems \cite{Hajek1967,Hajek1967a}.   

A NCMS is a triple $(T,Q,Tr)$, where $T$ is a time scale (which is assumed to be the real interval $[0,+\infty)$), $Q$ is a nonempty set (state space), and $Tr \subseteq (T \tilde\to Q)$ is a set of partial functions from $T$ to $Q$ which are called trajectories. To be a NCMS, by the definition the set $Tr$ must satisfy 3 rather non-restrictive properties (be closed under proper restrictions, be Markovian, and be complete in the sense that each nonempty chain of trajectories in the sense of the subtrajectory relation must have a supremum in $Tr$, see \cite{Ivanov2016} for the details). The mentioned Markovian property in this case is not formally related to the Markov processes in the probability theory, but has a rather similar idea: informally, it states that at any time moment, the possible future evolution of the state of the system depends on its current state, but not on the way by which the system reached it.

One can also view NCMS as a generalization of state transition systems
to continuous time, since the properties of $Tr$ stated in the definition of NCMS are satisfied for transition systems, when interpreted in terms of discrete time and transition system runs. By themselves NCMS describe evolutions of the state of a system (trajectories). If  necessary, it is also possible to associate  ``observable traces'' with trajectories (which are   quantities that evolve with trajectories or are simply pointwise projections of trajectories). After this association one gets an analog of a labeled transition system (LTS) in continuous time. We call this analog a labeled NCMS. This notion is defined in \cite{Ivanov2016}. 

Many concrete dynamical models  are instances of NCMS \cite{IvTAMC2014,Ivanov2012} (which are also examples of O. H\'{a}jek's solution systems), including models the behavior of which is described by classical and Car\'{a}theodory solutions of ordinary differential equations, ordinary solutions of differential inequalities and inclusions, switched systems, various hybrid dynamical systems.

The main feature of NCMS (not available for general O. H\'{a}jek's solution systems) is that they admit reduction of certain forms of analysis of their global-in-time behavior to analysis of their local-in-time behavior (which can be done using  methods depending on the means used to specify this behavior -- differential equations, inclusions, etc.; in all such cases the set of well-known local-in-time analysis methods is quite large, including  solution approximation methods with guaranteed error bounds, etc.). 
One
example
of this reduction is the method of proving the existence of global (i.e. defined on $T$) trajectories
which is based
on proving the existence of some  trajectories defined in a neighborhood of each time moment described in \cite{Ivanov2012}.

Like  in the case of other types of dynamical systems, one can study the behavior of a NCMS by analyzing its reach set.
A $t$-reach set is defined as the set of states which can be reached by a NCMS at the time moments in $[0,t]$ by following trajectories that  start at the time moment 0. Reach sets can have complex structure, so for system behavior analysis it is usually  more useful to consider their under- and over-approximations by well-behaved sets. One can consider this as a computational problem (compute under- and over-approximations of the $t$-reach set), or as a decision problem (check if a given candidate set is an under- or over-approximation of the $t$-reach set). 

In this paper we focus on the latter case, and, moreover, focus on methods of proving that its answer is affirmative. This view is useful when one considers applying  interactive proof assistants such as Mizar system, Isabelle, Coq, etc. to the verification of cyber-physical systems. 

 In this context, an obvious approach to proving that a given set $A$ is an overapproximation of a $t$-reach set is to show that it contains an invariant set (i.e. a set of states $I$ such that each trajectory that starts in $I$ remains forever in $I$) which includes the set of initial states of a NCMS.

Consider the problem of proving that a given set $A$ is an underapproximation of a $t$-reach set of a  NCMS.
Below we propose an approach to this problem based on finding a suitable superset $S$ of $A$ which has the property which, informally, means that that each partial trajectory of the system which lies entirely in $S$ either is defined as the initial time moment, or can be locally extended backward in time, or can be locally modified in such a way that the resulting trajectory can be locally extended back in time. This condition has a relatively simple logical expression and is convenient for applying a wide range of local existence theorems and local dynamics analysis methods  such as linearization of system's dynamics, various series expansions, approximations, singularity analysis to proving reachability. This  is important in the case of  using proof assistants for verifying bounds of  $t$-reach sets of continuous- or discrete-continuous models. For an example of formalization of NCMS and related results in the proof assistant Isabelle, we refer to \cite{Code}.
\section{Preliminaries}\label{sect_prelim}

We will use the following notation: $\mathbb{N}=\{1,2,3,...\}$ is the set of natural numbers; $\mathbb{R}$ is the set of real numbers, $\mathbb{R}_+$ is the set of nonnegative real numbers,  $f : A \to B$ is a total function from a set $A$ to  $B$; $f: A \tilde\to B$ is a partial function from $A$ to  $B$, $2^A$ is the power set of a set $A$, $f|_A$ is the restriction of a function $f$ to a set $A$.
For a function $f:A \tilde\to B$ we will use the notation $f(x) \downarrow$ ($f(x) \uparrow$) to denote that $f(x)$ is defined, or, respectively, is undefined on the argument $x$.

We will not distinguish  the notions of a function and a functional binary relation. When we write that a function $f:A \tilde\to B$  is total or surjective, we mean that $f$ is total on the set $A$ specifically ($f(x)$ is defined for all $x \in A$), or, respectively, is onto $B$ (for each $y \in B$  there exists $x \in A$  such that $y=f(x)$).

For any $f:A \tilde \to B$ denote $dom(f)=\{x\,|\,f(x)\downarrow\}$, $range(f)=\{y~|\exists x \in dom(f)~y=f(x)\}$. 
For any partial functions $f,g$ the notation $f(x)\cong g(x)$ will mean the strong equality: $f(x) \downarrow$ if and only if $g(x) \downarrow$,  and $f(x)\downarrow$ implies $f(x)=g(x)$.
We will denote by $f \circ g$ the functional composition: $(f \circ g) (x) \cong f(g(x))$ and by $f|_A$ the restriction of a function $f$ to a set $A$, i.e. a function defined on $dom(f) \cap A$ such that the graph of $f|_A$ is a subset of the graph of $f$.

Denote by $T$ the non-negative real time scale $[0,+\infty)$. We will assume that $T$ is equipped with a topology induced by the standard topology on $\mathbb{R}$. 

The symbols $\neg$, $\vee$, $\wedge$, $\Rightarrow$, $\Leftrightarrow$ will denote the logical operations of negation, disjunction, conjunction, implication, and equivalence respectively.

Let us denote by $\mathfrak{T}$ the set of all intervals  in $T$ (connected subsets) which have the cardinality greater than one. Let $Q$ be a set (a state space)  and $Tr$ be some set of functions of the form $s: A \to Q $, where $A \in \mathfrak{T}$. We will call the elements of $Tr$ \textit{(partial) trajectories}.
\begin{definition}[\cite{Ivanov2012,Ivanov2013new}]
A set of trajectories $Tr$ is closed under proper restrictions (CPR),  if  $s|_A \in Tr$ for each $s \in Tr$ and $A \in \mathfrak{T}$ such that $A \subseteq dom(s)$.
\end{definition}

Let us introduce the following notation: if $f,g$ are partial functions, $f \sqsubseteq g$ means that the graph of $f$ is a subset of the graph of $g$, and $f \sqsubset g$ means that the graph of $f$ is a proper subset of $g$.

\begin{definition} Let $s_1,s_2 \in Tr$ be trajectories. Then $s_1$ is called a subtrajectory of $s_2$, if $s_1 \sqsubseteq s_2$.
\end{definition}

The pair $(Tr, \sqsubseteq)$ is a possibly empty partially ordered set.

%\begin{definition}\label{def_detdsys} An abstract dynamical system $(T,Q,Tr)$ %is 
%\begin{itemize}
%\item[(1)]
%deterministic, if for each $\mathcal{T}_0 \in T$ and $q_0 \in Q$ the set 
%\[\{s \in Tr~|~s(\mathcal{T}_0)=q_0\}\] is a (possibly empty) chain with respect to % $\sqsubseteq$.
%\item[(2)]
%forward-deterministic,  if for each $\mathcal{T}_0 \in T$ and $q_0 \in Q$ the set % \[\{s|_{\{t\in T|t\ge \mathcal{T}_0 \}} ~|~ s\in Tr, s(\mathcal{T}_0)=q_0\}\] is a (possibly %empty) chain with respect to  $\sqsubseteq$.
%\end{itemize}

%\end{definition}

%Informally, (1) means that there is no more than one way to move  forward %or backward in time with respect to $\mathcal{T}_0$ from the state $q_0$ along some %trajectory and (2) is similar, but concerns only a forward-in-time movement.

\begin{definition}[\cite{Ivanov2012,Ivanov2013new}]\label{def_compl_mark_trset} A CPR set of trajectories $Tr$  is 
\begin{enumerate}
\item[(1)]
Markovian, if for each $s_1, s_2 \in Tr$ and  $t_0 \in T$ such that $t_0=\sup dom(s_1)= \inf dom(s_2)$, $s_1(t_{0})\downarrow$, $s_2(t_{0})\downarrow$, and $s_1(t_{0})=s_2(t_{0})$, the following function $s$ belongs to $Tr$:
$s(t) = s_1(t)$, if $t \in dom(s_1)$ and $s(t)=s_2(t)$, if $t \in dom(s_2)$.
\item[(2)]
complete, if each non-empty chain in $(Tr, \sqsubseteq)$ has a supremum.
\end{enumerate}
\end{definition}

\begin{definition}[\cite{Ivanov2012,Ivanov2013new}] A nondeterministic complete  Markovian system (NCMS) is a triple $(T, Q, Tr)$, where $Q$ is a set (called the state space) and $Tr$ (called the set of trajectories) is a set of functions $s: T \tilde\to Q$ such that $dom(s) \in \mathfrak{T}$, which is CPR, complete, and Markovian.
\end{definition}

Examples of representation of sets of trajectories of well-known continuous and discrete-continuous dynamical models such as ordinary differential equations, differential inclusions, switched systems, etc.  as can be found in \cite{Ivanov2016,Ivanov2012,Hajek1967,Hajek1967a}.

\section{Main Result}

\begin{definition}
Let $I_1, I_2 \in \mathfrak{T}$ and $s_1 : I_1 \to Q$, $s_2: I_2 \to Q$. Then $s_2$ is called a backward extension of $s_1$, if  $s_1 \sqsubset s_2$ and $t_2 \le t_1$ for each $(t_1, t_2) \in dom(s_1) \times dom(s_2)$.
\end{definition}

\begin{definition} Let $s_1 : I_1 \to Q$ where $I_1 \in \mathfrak{T}$.  

Then a  $\tau$-backward escape from $s_1$ at time $d$, where $d$ is a non-maximal element of $dom(s_1)$, is a function  $s_2: [c,d] \to Q$ for some $c \in T$, $c<d$ such that $\tau = d-c$ and $s_2(d) = s_1(d)$.
\end{definition}

\begin{definition} A trajectory $s \in Tr$ of a NCMS $(T,Q,Tr)$ is called
\begin{itemize}
\item initial, if $s(0)\downarrow$ (i.e. $s$ is defined at the initial time moment 0);
\item $S$-valued, where $S$ is a set, if $range(s) \subseteq S$.
\end{itemize}
\end{definition}

\begin{definition} Let $\Sigma=(T,Q,Tr)$ be a NCMS and $t_0 \in (0,+\infty)$.
\begin{itemize}
\item
The $t_0$-reach set of $\Sigma$ is the set 

$\{q \in Q~|~\exists s \in Tr_0 ~\exists t \in dom(s)\cap [0,t_0] ~q=s(t)\}$,

 where $Tr_0$ is the set of all initial trajectories of $\Sigma$ (i.e. the set of states that are attained at some time in $[0, t_0]$ by some initial trajectory).
\item
An underapproximation of a $t_0$-reach set of $\Sigma$ is a subset of the $t_0$-reach set of $\Sigma$.
\item
The $t_0$-right range set of $\Sigma$ is the set 

$\{q \in Q~|~\exists s \in Tr: dom(s) \subseteq [0,t_0] \wedge \max dom(s)\downarrow \wedge  q = s(\max dom(s)) \}$ 

(i.e. the set of states that are attainted at the right end of some trajectory defined within $[0,t_0]$, but not necessarily  an initial trajectory).
\item $\Sigma$ is $f$-backward extensible, where  $f : \mathbb{R}_+ \to \mathbb{R}_+$ is a function of class  $K$ (i.e. a continuous, strictly increasing function such that  $f(0)=0$), if for each  $s \in Tr$ at least one of the following  holds:
\begin{itemize}
\item[a)] $s$ is an initial trajectory;
\item[b)] there exists  $s' \in Tr$ such that $s'$ is a backward extension of $s$;
\item[c)]
$dom(s)$ has no minimum element and there exists $s' \in Tr$,  $t \in T$ -- a non-maximal element of $dom(s)$, and $\tau  \ge f(t - \inf dom(s))$ such that $s'$ is a $\tau$-backward escape from $s$ at time $t$.
\end{itemize}
\item
 A \textit{sub-NCMS of }$\Sigma$ is a NCMS $\Sigma=(T,Q',Tr')$ such that $Q \subseteq Q'$, $Tr \subseteq Tr'$.
\end{itemize}
\end{definition}

\begin{theorem}
Let $f : \mathbb{R}_+ \to \mathbb{R}_+$ be a function of class $K$, $\Sigma$ be a NCMS, $t_0 \in (0,+\infty)$, $A \subseteq Q$.

Then $A$ is an underapproximation of the $t_0$-reach set of $\Sigma$ \\if and only if  $A$ is a subset of the $t_0$-right range set of some $f$-backward extensible sub-NCMS of $\Sigma$.
\end{theorem}
\textit{Proof}. We will use the terminology of \cite{Ivanov2012,Ivanov2016,IvTAMC2014}. 
\begin{description}
\item[``If''] 
The idea is to apply the method of proving the existence of infinite-in-time trajectories of NCMS described in \cite{Ivanov2012}, but in reversed time and to show that locally defined trajectories that end in $A$ can be continued  backward all the way until the initial time moment using  left (time revered) \textit{ extensibility measures }\cite{Ivanov2016}.   Let $A$ be a subset of the $t_0$-right range set of $\Sigma'=(T,Q',Tr')$, where $\Sigma'$ is a $f$-backward extensible sub-NCMS of $\Sigma$.
Let $f^-(a,b)=a-f(a-b)$.  From the conditions of the $f$-backward extensible NCMS it  follows that $\Sigma'$ satisfies the local \textit{backward} extensibility property and each \textit{left} dead-end path in $\Sigma'$ is $f^-$-{escapable}. Then by the  theorem about the left-dead end path (a time reversal of the theorem about the right-dead end path as formulated in \cite{Ivanov2016}), each left dead-end path in $\Sigma'$ is strongly escapable.

 Let $q \in A$. Then $q = s(\max dom(s))$ for some $s \in Tr'$ such that $dom(s) \subseteq [0,t_0]$.  Denote $b = \max dom(s)$. Then $s \in Tr$. If $s$ is an initial trajectory or is a subtrajectory of an initial trajectory of $\Sigma'$, then $q$ is attained by an initial trajectory of $\Sigma$ at a time in $[0,t_0]$, so $q$  is in the $t_0$-reach set of $\Sigma$. 

Assume the contrary, i.e. $s$ is not a subtrajectory of an initial trajectory of $\Sigma'$. Suppose that there exists a trajectory of $\Sigma'$ which is a backward extension of $s$. Then the completeness property of NCMS implies that $Tr'$ contains a maximal (in the sense of the subtrajectory relation $\sqsubseteq$) backward extension of $s$ (among backward extensions of $s$ in $Tr'$).  Denote it as $s'$. Then $s'$ cannot be an initial trajectory, since otherwise $s$ would be a subtrajectory of an initial trajectory of $\Sigma'$. Also, $s'$ has no backward extension in $Tr'$. Since $\Sigma'$ is $f$-backward extensible, $dom(s')$ does not have a minimal element. Then $dom(s')=(a,b]$ for some $a \in T$ such that $a<b$. Then $s'$ is a \textit{left dead-end path} and it is \textit{strongly escapable}. Moreover, $q=s'(b)$ and $b \le t_0$. This implies that  $q$ is attained by an initial trajectory of $\Sigma$ at some time in $[0,t_0]$, so $q$ belongs to the $t_0$-reach set of $\Sigma$.  

Suppose that there is no trajectory of $\Sigma'$ which is a backward extension of $s$. Also $s$ is not an initial trajectory. Since $\Sigma'$ is $f$-backward extensible, $dom(s)$ does not  have a minimal element. Then $s$ is a \textit{left dead-end path} in $\Sigma'$ and it is \textit{strongly escapable}. Moreover, $q=s(b)$ and $b \le t_0$. This implies that  $q$ is attained by an initial trajectory of $\Sigma$ at some time in $[0,t_0]$, so $q$ is in the $t_0$-reach set of $\Sigma$.
 
\item[``Only if''] Let  $A$ be a subset of the $t_0$-reach set of $\Sigma$. Let $Tr_0$ be the set of initial trajectories of $\Sigma$, $Tr'=\{s': I \to Q ~|~ I \in \mathfrak{T} \wedge \exists s \in Tr_0~s'=s|_I\}$, and $\Sigma'=(T,Q,Tr')$. It is easy to see that $Tr'$ is CPR, Markovian, and complete (in the sense of Definition 3), so $\Sigma'$ is a NCMS. Moreover, it is a sub-NCMS of $\Sigma$, and is $f$-backward extensible, since each non-initial trajectory of $\Sigma'$  has a backward extension in $Tr'$. Moreover, for each initial trajectory $s$ of $\Sigma$, $s|_{[0,t_0]}$ is a trajectory with the domain in $[0,t_0]$ and by the CPR property of the NCMS,  its range is attainted at the right ends of  trajectories of $\Sigma'$ defined within $[0,t_0]$, so $A$ is a subset of the $t_0$-right range set of $\Sigma'$.  Thus $A$ is a subset of the $t_0$-right range set of $\Sigma'$, a $f$-backward extensible sub-NCMS of $\Sigma$.
\end{description}

An obvious way of obtaining a sub-NCMS of a NCMS $\Sigma$ is restricting the state space of $\Sigma$. More specifically, it is easy to check that if $S \subseteq Q$ and $Tr_S$ is the set of all $S$-valued trajectories of $\Sigma$, then $\Sigma_S = (T,S,Tr_S)$ is a sub-NCMS of $\Sigma$.
This implies the  above-mentioned approach to proving that  a given set $A$ is an underapproximation of a $t_0$-reach set of a  NCMS: to prove this it is sufficient to choose a function $f$ of class $K$ and find/guess a superset $S$ of $A$ which has the property that  $A$ is a subset of the $t_0$-right range set of $\Sigma_S$ and $\Sigma_S$ is $f$-backward extensible.

%\nocite{*}
\bibliographystyle{eptcs}
\bibliography{bib}
\end{document}